\documentclass[twocolumn,numberedappendix]{openjournal}
\usepackage{graphicx,amsmath,amssymb,amstext}
\usepackage{amsbsy,amsfonts,amsthm,color}
\usepackage[colorlinks,linkcolor=blue,citecolor=blue,urlcolor=blue ]{hyperref}
\usepackage[utf8]{inputenc}
\usepackage{float}
\usepackage{xcolor}
\usepackage{ulem}
\usepackage[T1]{fontenc}
\usepackage[title]{appendix}

\begin{document}

\title{Forecasting the accuracy of velocity-field reconstruction\vspace{-4em}}

\author{C. Blake$^{1,*}$}
\author{R.J. Turner$^1$}
\thanks{$^*$E-mail: cblake@swin.edu.au}
\affiliation{$^1$ Centre for Astrophysics and Supercomputing, Swinburne University of Technology, P.O. Box 218, Hawthorn, VIC 3122, Australia \\
}

\begin{abstract}
Joint analyses of the large-scale distribution of galaxies, and their motions under the gravitational influence of this density field, allow powerful tests of the cosmological model, including measurement of the growth rate of cosmic structure.  In this paper we perform a statistical comparison between two important classes of method for performing these tests.  In the first method, which we refer to as the ``power-spectrum method'', we measure the 2-point power spectra between the velocity and density tracers, and jointly fit these statistics using theoretical models.  In the second method, which we refer to as the ``reconstruction-and-scaling method'', we use the density tracers to reconstruct a model velocity field through space, which we compare with the measured galaxy velocities on a point-by-point basis.  By generating an ensemble of numerical simulations in a simplified test scenario, we show that the error in the growth rate inferred by the reconstruction-and-scaling method may be under-estimated, unless the full covariances of the underlying and reconstructed velocity fields are included in the analysis.  In this case the inferred growth rate errors agree with both the power-spectrum method and a Fisher matrix forecast.  We provide a roadmap for evaluating these covariances, considering reconstruction performed using both a Fourier basis within a cuboid, and a Spherical Fourier-Bessel basis within a curved-sky observational volume.
  \\[1em]
\end{abstract}

\maketitle

\section{Introduction}

Direct measurements of the peculiar velocities of galaxies constitute a powerful dataset for testing the cosmological model.  The correlation structure in the velocity field, which is organised in large-scale coherent flows, probes the growth rate of cosmic structure encoding gravitational physics \citep[general cosmological forecasts of such measurements are provided by, e.g.,][]{2014MNRAS.445.4267K, 2017MNRAS.464.2517H}.  The utility of the method arises from the close connection of the velocity and density fields in linear theory, the sensitivity of velocities to density modes on large scales, and the (partial) cancellation that occurs between the sample variance imprinted in the fields \citep{2009JCAP...10..007M}.  The radial peculiar velocities of galaxies may be directly determined by combining redshift measurements with independent distance indicators such as supernovae, the Fundamental Plane of early-type galaxies, or the Tully-Fisher relation for late-type galaxies \citep[for a historical review, see][]{1995PhR...261..271S}.

Several methodologies exist for connecting the measured peculiar velocities of galaxies to the underlying density field in order to measure the growth rate of structure.  First, two-point correlations between the velocity and density tracers may be measured, either as a configuration-space correlation function \citep[e.g.,][]{1988ApJ...332L...7G, 2000Sci...287..109J, 2018MNRAS.480.5332W, 2019MNRAS.486..440D, 2023MNRAS.518.2436T, 2024MNRAS.532.3972L} or as a Fourier-space power spectrum \citep[e.g.,][]{2000MNRAS.319..573P, 2017MNRAS.471.3135H, 2019MNRAS.487.5235Q} -- which we call the ``power-spectrum method''.  Second, the measured galaxy density field may be used to ``reconstruct'' a model velocity field through space, which may then be compared with the measured peculiar velocities on a point-by-point basis \citep[e.g.,][]{2005ApJ...635...11P, 2012MNRAS.420..447T, 2015MNRAS.450..317C, 2020MNRAS.497.1275S, 2020MNRAS.498.2703B, 2021MNRAS.507.1557L, 2024MNRAS.531...84B} -- which we call the ``reconstruction-and-scaling method''.  Third, density and velocity fields may be constructed on grids through space, and their mutual correlations analysed using maximum-likelihood methods \citep[e.g.,][]{2014MNRAS.444.3926J, 2017JCAP...05..015H, 2020MNRAS.494.3275A, 2023MNRAS.518.1840L}.

Multiple techniques exist for reconstructing the velocity field from the observed galaxy distribution \citep[with examples of early work including][]{1989ApJ...336L...5B, 1991ApJ...372..380Y, 1993ApJ...412....1D, 1994ApJ...421L...1N, 1994MNRAS.266..475H, 1999ApJ...520..413Z}.  Relying on linear theory alone, the velocity components may be constructed from the density field on a Cartesian grid using Fourier transforms (see Eq.\ref{eq:uk} below), similar to the techniques that are employed when computing displacements for ``reconstructing'' the baryon acoustic peak \citep{2007ApJ...664..675E, 2015MNRAS.453..456B}.  Alternatively, within a spherical geometry that may be better-matched to realistic observational window functions, the density field may be expressed as coefficients of a Spherical Fourier-Bessel (SFB) transform \citep[e.g.,][]{1995MNRAS.272..885F, 1995MNRAS.275..483H, 2006MNRAS.373...45E, 2012A&A...540A..60L, 2014PhRvD..90f3515N, 2021MNRAS.507.1557L, 2022arXiv221205760K, 2024PhRvD.109h3502G}.  A rich set of other methods have been proposed to access information beyond linear theory, including sophisticated Bayesian physical models \citep[e.g.,][]{2012MNRAS.427L..35K, 2013MNRAS.432..894J, 2016MNRAS.457..172L} and convolutional neural networks \citep[e.g.,][]{2023JCAP...06..062Q}.

Datasets such as the 6-degree Field Galaxy Survey peculiar velocity sample \citep{2012MNRAS.427..245M, 2014MNRAS.445.2677S}, the Sloan Digital Sky Survey sample \citep{2022MNRAS.515..953H}, or compilations such as \textit{CosmicFlows} \citep{2016AJ....152...50T, 2023ApJ...944...94T}, in conjunction with overlapping galaxy redshift survey data, have allowed measurements of the local growth rate to be performed using many of these techniques.  Current measurements may be summarised as follows \citep[see also Fig.7 of][]{2023MNRAS.518.2436T}: methods based on the 2-point correlation structure of the fields have recovered the local growth rate with $15 - 20\%$ accuracy \citep[e.g.,][]{2014MNRAS.444.3926J, 2019MNRAS.486..440D, 2019MNRAS.487.5235Q, 2020MNRAS.494.3275A, 2021MNRAS.507.1557L, 2023MNRAS.518.2436T}, and the measured values agree with the predictions of standard cosmological models \citep{2020A&A...641A...6P}.  However, growth rate measurements involving the reconstruction-and-scaling method have typically yielded significantly smaller $4 - 8\%$ errors \citep[e.g.,][]{2015MNRAS.450..317C, 2020MNRAS.497.1275S, 2020MNRAS.498.2703B, 2024MNRAS.531...84B}, accompanied by some suggestions of discrepancies with standard model predictions \citep{2020MNRAS.497.1275S, 2024MNRAS.531...84B}.

This potential difference between growth rate measurement errors, when applying different techniques to comparable datasets, motivated us to perform a detailed comparison between the power-spectrum and reconstruction-and-scaling analysis methodologies, using a consistent set of numerical simulations.  We pay particularly close attention to modelling the covariance of the reconstructed velocity field, which arises from sample variance and noise, and propagating this error into the local comparison with a measured peculiar velocity.  We study these effects using both Fourier-based reconstruction within a cube (with a flat-sky approximation considering the velocity component along a single axis) and a Spherical Fourier-Bessel reconstruction within a sphere (using a curved-sky treatment in which our observable is the radial velocity).  We base our analysis on linear-theory reconstruction methodologies because these allow convenient analytical treatments of velocity correlations, which we can self-consistently include in maximum-likelihood analyses.

Our paper is structured as follows: in Sec.\ref{sec:model} we outline our models for the correlations of the underlying velocity field, the reconstructed velocity field in the Fourier and SFB basis, and the cross-correlations between the two fields.  In Sec.\ref{sec:method} we introduce the numerical simulations we use to test the different methodologies, describe our implementation of the reconstruction-and-scaling method, and overview our 2-point power spectrum analysis.  In Sec.\ref{sec:results} we compare the growth rate fits using the different techniques, and we conclude in Sec.\ref{sec:conc}.

\section{Modelling velocity correlations}
\label{sec:model}

In this section we derive models for the correlation of the velocity field between two locations $\mathbf{x}$ and $\mathbf{y}$, which are needed to compute the likelihood of growth rate fits based on reconstruction. We distinguish two different velocity fields in our analysis.  First, we consider the underlying ``true'' velocity field, which we write as $\mathrm{u}_i(\mathbf{x})$ (for a component $i = \{ x,y,z \}$) and $\mathrm{u}_r(\mathbf{x})$ (for the radial velocity).  The true velocity field may be estimated using direct galaxy peculiar velocity measurements.  Second, we consider the reconstructed velocity field as inferred from the measured galaxy density field, which we correspondingly write as $\mathrm{v}_i(\mathbf{x})$ or $\mathrm{v}_r(\mathbf{x})$.  We will consider velocity reconstructions performed in both a Fourier basis (appropriate for analysis in a cube) and a Spherical Fourier-Bessel basis (appropriate for analysis within a sphere).

\subsection{The underlying velocity field}
\label{sec:corru}

We base our analysis in linear theory of the growth of structure, which allows us to create analytical expressions for the velocity correlations.  The relation between the underlying velocity components and matter overdensity field in linear theory is \citep[for a derivation see e.g., Appendix C of][]{2017MNRAS.471..839A},
\begin{equation}
  \tilde{\mathrm{u}}_i(\mathbf{k}) = -iaHf \frac{k_i}{k^2} \, \tilde{\delta}_m(\mathbf{k}) ,
\label{eq:uk}
\end{equation}
where $\tilde{\mathrm{u}}_i(\mathbf{k})$ is the Fourier transform of the $i = \{ x,y,z \}$ component of the velocity field as a function of wavevector $\mathbf{k}$, $\tilde{\delta}_m(\mathbf{k})$ is the Fourier transform of the matter overdensity field, $a$ is the cosmic scale factor, $H$ is the Hubble parameter and $f$ is the growth rate of cosmic structure.  Inverse Fourier-transforming Eq.\ref{eq:uk} we obtain,
\begin{equation}
  \mathrm{u}_i(\mathbf{x}) = -iaHf \int \frac{V_\mathrm{box} \, d^3\mathbf{k}}{(2\pi)^3} \, \frac{k_i}{k^2} \, \tilde{\delta}_m(\mathbf{k}) \, e^{-i\mathbf{k} \cdot \mathbf{x}} ,
\label{eq:ui}
\end{equation}
where $V_\mathrm{box}$ is the volume of the enclosing Fourier cuboid, and we will neglect redshift-space distortions throughout this study.  The correlation of velocity components between two locations $\mathbf{x}$ and $\mathbf{y}$ can then be derived as \citep{1988ApJ...332L...7G},
\begin{equation}
    \langle \mathrm{u}_i(\mathbf{x}) \, \mathrm{u}_j(\mathbf{y}) \rangle \\
    = a^2 H^2 f^2 \int \frac{d^3\mathbf{k}}{(2\pi)^3} \frac{k_i \, k_j}{k^4} P_m(\mathbf{k}) e^{-i\mathbf{k} \cdot \mathbf{r}} ,
\label{eq:uiujcov}
\end{equation}
where $P_m(k)$ is the matter power spectrum and the vector separation of the locations is $\mathbf{r} = \mathbf{x} - \mathbf{y}$.  We use the notation $\langle ... \rangle$ to indicate an average over many cosmological realisations.

The radial velocity at a location relative to an observer at the origin is given in terms of these velocity components as $\mathrm{u}_r(\mathbf{x}) = \sum_i \mathrm{u}_i(\mathbf{x}) \, x_i/x$.    Using Eq.\ref{eq:uiujcov} we can derive the correlation between the radial velocities at two locations \citep[e.g.,][]{1988ApJ...332L...7G, 2024MNRAS.527..501B},
\begin{equation}
\begin{split}
    \langle \mathrm{u}_r(\mathbf{x}) \, & \mathrm{u}_r(\mathbf{y}) \rangle = a^2 H^2 f^2 \int \frac{dk}{2\pi^2} \, P_m(k) \\ &\times \left[ \frac{j_1(kr)}{kr} \left( \hat{\mathbf{x}} \cdot \hat{\mathbf{y}} \right) - j_2(kr) \left( \hat{\mathbf{x}} \cdot \hat{\mathbf{r}} \right) \left( \hat{\mathbf{y}} \cdot \hat{\mathbf{r}} \right) \right] ,
\end{split}
\label{eq:ururcov}
\end{equation}
where $j_\ell$ are the spherical Bessel functions of order $\ell$.  The variance of the underlying velocity at a point is then,
\begin{equation}
    \langle \mathrm{u}^2(\mathbf{x}) \rangle = \frac{a^2 H^2 f^2}{6\pi^2} \int dk \, P_m(k) ,
\label{eq:siguu}
\end{equation}
which may be applied to the radial velocity or any component.  We assume that we can use velocity tracers with redshift-independent distance determinations to perform unbiased measurements of these underlying velocities with standard deviation $\epsilon_\mathrm{u}(\mathbf{x})$.  When including this measurement noise, a term $\epsilon_\mathrm{u}^2$ must be added to Eq.\ref{eq:siguu}.

\subsection{Fourier velocity reconstruction}
\label{sec:corrvfour}

Whilst the matter overdensity field $\delta_m(\mathbf{x})$ in Eq.\ref{eq:uk} is unknown, it may be estimated from the observed galaxy overdensity field $\delta_g(\mathbf{x})$, enabling an approximate reconstruction of the velocity field to be performed.  We will assume throughout this study that the matter and galaxy overdensity fields can be related by a linear bias factor $b$, such that the galaxy power spectrum is given by $P_g(k) = b^2 \, P_m(k)$.

We suppose this density field estimate is carried out using $N_g$ galaxies within the Fourier cuboid with average number density $n_0 = N_g/V_\mathrm{box}$, where the observed galaxy number density field is $n_g(\mathbf{x})$, and the galaxies have been Poisson-sampled from a known window function $W(\mathbf{x}) = \langle n_g(\mathbf{x}) \rangle$.  Following the standard approach in power spectrum estimation, we define the galaxy overdensity as,
\begin{equation}
  \delta_g(\mathbf{x}) = \frac{n_g(\mathbf{x}) - W(\mathbf{x})}{n_0} ,
\end{equation}
where the alternative form $\delta_g(\mathbf{x}) = n_g(\mathbf{x})/W(\mathbf{x}) - 1$ is avoided because it can involve dividing by small quantities.  We can then show that \citep[e.g.,][]{1994ApJ...426...23F, 2013MNRAS.436.3089B},
\begin{equation}
\begin{split}
  &\langle \tilde{\delta}_g(\mathbf{k}) \, \tilde{\delta}_g^*(\mathbf{k}') \rangle = \frac{1}{n_0} \tilde{W}(\mathbf{k}-\mathbf{k}') \\ & + \frac{b^2}{n_0^2} \int \frac{d^3\mathbf{k}''}{(2\pi)^3} \, P_m(\mathbf{k}'') \, \tilde{W}(\mathbf{k}'-\mathbf{k}'') \, \tilde{W}^*(\mathbf{k}-\mathbf{k}'') ,
\end{split}
\label{eq:dgdg}
\end{equation}
where $\tilde{W}(\mathbf{k})$ is the Fourier transform of the window function. Eq.\ref{eq:dgdg} includes a contribution from Poisson noise (represented by the first term) and sample variance (represented by the second term, which involves a convolution of the power spectrum and window function).

We can now determine the statistics of the reconstructed velocity field components $\mathrm{v}_i(\mathbf{x})$, or the radial field $\mathrm{v}_r(\mathbf{x})$, by applying Eq.\ref{eq:ui} using $\delta_g(\mathbf{x})/b$ in place of $\delta_m(\mathbf{x})$.  We note that in a reconstruction algorithm, the measured density field is sometimes smoothed by a Gaussian filter with standard deviation $\lambda$, to reduce noise.  In this case the underlying power spectra of the sample variance and noise are damped by a factor $D^2(k)$, where $D(k) = e^{-k^2 \lambda^2/2}$ is the Fourier transform of the damping kernel.

For the purpose of this analysis we will assume a uniform selection function $W(\mathbf{x}) = n_0$ (we provide an approximate treatment of the more general case in Appendix \ref{sec:approxcoeff}).  In this case we find that the covariance of the reconstructed velocities between two positions, $\langle \mathrm{v}_i(\mathbf{x}) \, \mathrm{v}_j(\mathbf{y}) \rangle$ or $\langle \mathrm{v}_r(\mathbf{x}) \, \mathrm{v}_r(\mathbf{y}) \rangle$, is obtained using the same relations given in Eq.\ref{eq:uiujcov} and Eq.\ref{eq:ururcov} whilst making the replacement,
\begin{equation}
    P_m(k) \rightarrow \left( P_m(k) + \frac{1}{n_0} \right) D^2(k) .
\end{equation}
The variance of the reconstructed velocity at a point is hence given by,
\begin{equation}
    \langle \mathrm{v}^2(\mathbf{x}) \rangle = \frac{a^2 H^2 f^2}{6\pi^2} \int dk \, \left( P_m(k) + \frac{1}{n_0} \right) D^2(k) ,
\label{eq:sigvv}
\end{equation}
which may be compared with the equivalent result for the underlying velocity field shown in Eq.\ref{eq:siguu}.

The cross-correlation between the underlying velocity field and the reconstructed velocity field, $\langle \mathrm{u}_i(\mathbf{x}) \, \mathrm{v}_j(\mathbf{y}) \rangle$ or $\langle \mathrm{u}_r(\mathbf{x}) \, \mathrm{v}_r(\mathbf{y}) \rangle$, may be determined using,
\begin{equation}
  \langle \tilde{\delta}_m(\mathbf{k}) \, \tilde{\delta}_g^*(\mathbf{k}') \rangle = \frac{b}{n_0} P_m(\mathbf{k}) \, \tilde{W}(\mathbf{k}-\mathbf{k}') .
\end{equation}
Using similar approximations as above, the cross-correlation is obtained by making the replacement in Eq.\ref{eq:uiujcov} and Eq.\ref{eq:ururcov},
\begin{equation}
    P_m(k) \rightarrow P_m(k) \, D(k) ,
\end{equation}
and the covariance of the velocity fields at a point is given by,
\begin{equation}
    \langle \mathrm{u}(\mathbf{x}) \, \mathrm{v}(\mathbf{x}) \rangle = \frac{a^2 H^2 f^2}{6\pi^2} \int dk \, P_m(k) \, D(k) .
\label{eq:siguv}
\end{equation}
These relations are also discussed in \cite{2023MNRAS.526..337T}.

\subsection{Spherical Fourier-Bessel velocity reconstruction}
\label{sec:corrvsfb}

Spherical Fourier-Bessel (SFB) transforms constitute a natural basis for decomposing density fluctuations within a spherical volume representing a curved-sky window function \citep[e.g.,][]{1995MNRAS.272..885F, 1995MNRAS.275..483H, 2006MNRAS.373...45E, 2012A&A...540A..60L, 2014PhRvD..90f3515N, 2021MNRAS.507.1557L, 2022arXiv221205760K, 2024PhRvD.109h3502G}.  We refer to \cite{1995MNRAS.272..885F} for a full description of the methodology of an SFB velocity reconstruction analysis, providing a brief overview here.

In an SFB expansion of an overdensity field $\delta(\mathbf{x})$ within a spherical volume bounded by $|\mathbf{x}| = R$, we create a linear superposition of the basis functions $j_\ell(k_{n \ell} x) \, Y_{\ell m}(\hat{\mathbf{x}})$, where $k_{n \ell}$ are discrete wavenumbers defined below, $Y_{\ell m}$ are spherical harmonic functions, and the integers $\{ \ell, m, n \}$ index the modes.  These basis functions are combined with amplitudes $\delta_{\ell m n}$ and normalisations $c_{n \ell}$:
\begin{equation}
    \delta(\mathbf{x}) = \sum_{\ell m n} c_{n \ell} \, j_\ell(k_{n \ell} x) \, Y_{\ell m}(\hat{\mathbf{x}}) \, \delta_{\ell m n} .
\label{eq:sfb}
\end{equation}
The sum over modes may be written out more fully as,
\begin{equation}
    \sum_{\ell m n} \rightarrow \sum_{n=0}^{n_\mathrm{max}} \sum_{\ell=0}^{\ell_\mathrm{max}(n)} \sum_{m = -\ell}^{+\ell} ,
\end{equation}
illustrating that in a practical case the sums are limited to $n \le n_\mathrm{max}$ and $\ell \le \ell_\mathrm{max}$, representing the radial and angular resolution of the expansion, respectively.  In our study we select these limits such that $k_{n \ell} < k_\mathrm{max}$, where $k_\mathrm{max}$ is a specified resolution scale.

In the discretisation of SFB modes represented by Eq.\ref{eq:sfb}, boundary conditions have been applied at $|\mathbf{x}| = R$.  We assume in our treatment that the logarithmic derivative of the gravitational potential is continuous at this boundary (see \cite{1995MNRAS.272..885F} for further discussion of this boundary condition and other possible choices), which leads to a condition on the discrete wavenumbers,
\begin{equation}
    j_{\ell-1}(k_{n \ell} R) = 0 .
\end{equation}
We note that for $\ell = 0$, recursion relations allow us to define $j_{-1}(u) = \cos{u}/u$.  For this boundary condition, the normalisation of the orthogonality relations of spherical Bessel functions leads to \citep{1995MNRAS.272..885F},
\begin{equation}
    c^{-1}_{n \ell} = \frac{1}{2} \, R^3 \, \left[ j_\ell(k_{n \ell} R) \right]^2 .
\end{equation}
The inverse transform of Eq.\ref{eq:sfb} is,
\begin{equation}
    \delta_{\ell m n} = \int_0^R dx \, x^2 \, j_\ell(k_{n \ell} x) \int d\Omega_x \, Y^*_{\ell m}(\hat{\mathbf{x}}) \, \delta(\mathbf{x}) ,
\label{eq:denstosfb}
\end{equation}
where similar to the previous section, we have assumed a uniform window function within the volume $|\mathbf{x}| < R$.  In this case the SFB coefficients are related to the underlying power spectrum as \citep{1995MNRAS.272..885F},
\begin{equation}
    \langle \delta_{\ell m n} \, \delta^*_{\ell' m' n'} \rangle = C_{n n' \ell} \, \delta^K_{\ell \ell'} \, \delta^K_{m m'} ,
\label{eq:powsfb}
\end{equation}
where $\delta_{ij}^K$ is the Kronecker delta and,
\begin{equation}
    C_{n n' \ell} = \frac{2}{\pi} \int dk \, k^2 \, W_{n \ell}(k) \,  W_{n' \ell}(k) \, P_m(k) ,
\end{equation}
in terms of the window function,
\begin{equation}
    W_{n \ell}(k) = \int_0^R dx \, x^2 \, j_\ell(k_{n \ell} x) \, j_\ell(kx) .
\end{equation}

Given the SFB coefficients of the density field, the reconstructed radial velocity field can be derived as \citep{1995MNRAS.272..885F},
\begin{equation}
    \mathrm{v}_r(\mathbf{x}) = aHf \sum_{\ell m n} c_{n \ell} \, k^{-1}_{n \ell} \, j'_\ell(k_{n \ell} x) \, Y_{\ell m}(\hat{\mathbf{x}}) \, \delta_{\ell m n} ,
\label{eq:vrecsfb}
\end{equation}
where $j'_\ell(u) = dj_\ell(u)/du$ represents the derivative of the spherical Bessel functions with respect to the argument.  Combining Eq.\ref{eq:vrecsfb} and Eq.\ref{eq:powsfb} allows us to derive the covariance of the reconstructed velocity field,
\begin{equation}
\begin{split}
    \langle \mathrm{v}_r(\mathbf{x}) \, & \mathrm{v}_r(\mathbf{y}) \rangle = a^2 H^2 f^2 \sum_{n n' \ell} k^{-1}_{n \ell} \, k^{-1}_{n' \ell} \, C_{n n' \ell} \\ &\times \left( \frac{2\ell+1}{4\pi} \right) j'_\ell(k_{n \ell} x) \, j'_\ell(k_{n' \ell} y) \, L_\ell(\hat{\mathbf{x}} \cdot \hat{\mathbf{y}}) .
\end{split}
\label{eq:vrvrcovsfb}
\end{equation}
In Appendix \ref{sec:sfbtocont} we demonstrate that Eq.\ref{eq:vrvrcovsfb} reduces to the correlation of the underlying radial velocity field, i.e.\ Eq.\ref{eq:ururcov}, in the limit that $R \rightarrow \infty$.

To express the cross-correlation between the underlying and reconstructed velocity field in the SFB basis, we use the expression for Eq.\ref{eq:vrecsfb} in the limit that $R \rightarrow \infty$,
\begin{equation}
    \mathrm{u}_r(\mathbf{x}) = aHf \int dk \sum_{\ell m} \sqrt{\frac{2}{\pi}} j'_\ell(kx) \, Y_{\ell m}(\hat{\mathbf{x}}) \, \delta_{\ell m}(k) ,
\end{equation}
where $\delta_{\ell m}(k)$ is the continuous spherical Fourier-Bessel transform of the overdensity field,
\begin{equation}
    \delta_{\ell m}(k) = \sqrt{\frac{2}{\pi}} k \int d^3\mathbf{x} \, \delta(\mathbf{x}) \, j_\ell(kx) \, Y^*_{\ell m}(\hat{\mathbf{x}}) .
\end{equation}
The discrete coefficients are related to the continuous transform as \citep[e.g.,][]{2022arXiv221205760K},
\begin{equation}
    \delta_{\ell m n} = \sqrt{\frac{2}{\pi}} \int dk \, k \, W_{n \ell}(k) \, \delta_{\ell m}(k) ,
\end{equation}
where,
\begin{equation}
  \langle \delta_{\ell m}(k) \, \delta^*_{\ell' m'}(k') \rangle = P_m(k) \, \delta_D(k-k') \, \delta^K_{\ell \ell'} \, \delta^K_{m m'} ,
\end{equation}
where $\delta_D$ is the Dirac delta function.  The cross-correlation function is hence given by,
\begin{equation}
\begin{split}
  \langle \mathrm{u}_r(\mathbf{x}) \, & \mathrm{v}_r(\mathbf{y}) \rangle = a^2 H^2 f^2 \sum_{n \ell} k^{-1}_{n \ell} \, j'_\ell(k_{n \ell} y) \, (2\ell+1) \\ & \times L_\ell(\hat{\mathbf{x}} \cdot \hat{\mathbf{y}}) \int \frac{dk \, k}{2\pi^2} \, W_{n\ell}(k) \, j'_\ell(kx) \, P_m(k) .
\end{split}
\label{eq:urvrcovsfb}
\end{equation}
In Sec.\ref{sec:method} below, we will validate these analytical expressions for the velocity correlation functions using numerical simulations.

\subsection{On the limitations of the correlation models}

We emphasise that the velocity correlation models derived in Sec.\ref{sec:corru}, Sec.\ref{sec:corrvfour} and Sec.\ref{sec:corrvsfb} contain approximations implying they cannot currently be applied to real datasets: they assume a uniform window function (within a box or sphere), linear growth theory, linear galaxy bias and they neglect redshift-space distortions.  We will use these correlation models to compare different methods for measuring the growth rate in this simplified test scenario.

\subsection{Bias in the reconstructed velocity}
\label{sec:bias}

The underlying and reconstructed velocity fields in the linear theory approximation are correlated Gaussian fields.  At any location $\mathbf{x}$ their joint statistics are well-described by an ``elliptical Gaussian'' 2D statistical relation between the model velocity $\mathrm{v}$ and underlying (or measured) velocity $\mathrm{u}$ \citep{2023MNRAS.526..337T}:
\begin{equation}
  P(\mathrm{u},\mathrm{v}) \propto \exp{\left[ - \frac{1}{2(1-r^2)} \left( \frac{\mathrm{u}^2}{\sigma_\mathrm{u}^2} - \frac{2 r \mathrm{u} \mathrm{v}}{\sigma_\mathrm{u} \sigma_\mathrm{v}} + \frac{\mathrm{v}^2}{\sigma_\mathrm{v}^2} \right) \right]} ,
\end{equation}
in terms of $\sigma_\mathrm{v}^2 = \langle \mathrm{v}^2(\mathbf{x}) \rangle$, $\sigma_\mathrm{u}^2 = \langle \mathrm{u}^2(\mathbf{x}) \rangle$, and the cross-correlation coefficient $r = \sigma_{\mathrm{u} \mathrm{v}}^2/(\sigma_\mathrm{u} \, \sigma_\mathrm{v})$, where $\sigma_{\mathrm{u} \mathrm{v}}^2 = \langle \mathrm{u}(\mathbf{x}) \, \mathrm{v}(\mathbf{x}) \rangle$.

\begin{figure}
\centering
\includegraphics[width=\columnwidth]{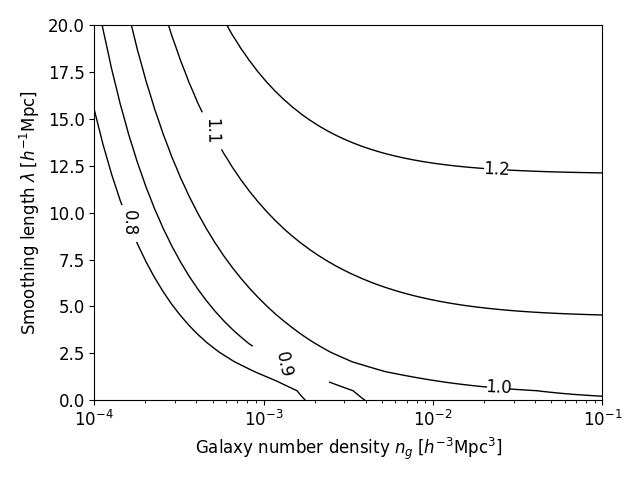}
\caption{The reconstructed velocity $\mathrm{v}$ at any location contains a multiplicative bias, $\alpha$, such that the mean measured velocity is $\overline{\mathrm{u}} = \alpha \, \mathrm{v}$.  This figure shows contours of $\alpha$ as a function of the galaxy number density used for the reconstruction, $n_g$, and the Gaussian smoothing scale, $\lambda$, applied to the number density distribution, assuming linear theory and a Fourier-based reconstruction with no window function.  Contours are shown corresponding to $\alpha = \{ 0.8, 0.9, 1.0, 1.1, 1.2 \}$.}
\label{fig:velbias}
\end{figure}

Given this relation, the velocity field model at any location contains a multiplicative bias \citep{2023MNRAS.526..337T} such that the mean measured value $\overline{\mathrm{u}}$, given a predicted value $\mathrm{v}$, is,
\begin{equation}
  \overline{\mathrm{u}} = \int_{-\infty}^\infty P(\mathrm{u},\mathrm{v}) \, d\mathrm{u} = \frac{r \, \sigma_\mathrm{u}}{\sigma_\mathrm{v}} \, \mathrm{v} \ne \mathrm{v} .
\end{equation}
Fig.\ref{fig:velbias} displays the magnitude of this multiplicative bias $\alpha = r \, \sigma_\mathrm{u}/\sigma_\mathrm{v}$ as a function of the number density of the galaxy survey used to perform the reconstruction, $n_g$, and the smoothing length, $\lambda$.  The bias is independent of the measurement noise $\epsilon_\mathrm{u}$.  In this calculation we have assumed the Fourier velocity reconstruction with no window function, and used Eq.\ref{eq:siguu}, Eq.\ref{eq:sigvv} and Eq.\ref{eq:siguv} to determine the variances $\langle \mathrm{u}^2 \rangle$, $\langle \mathrm{v}^2 \rangle$ and $\langle \mathrm{u} \mathrm{v} \rangle$.  We note that the reconstructed velocity becomes unbiased ($\alpha \rightarrow 1$) in the limit of high number density and no smoothing\footnote{As discussed by \cite{2021MNRAS.502.3723H}, the linear theory relation between velocity and density is modified at small scales by a damping term \citep[e.g.][]{2013PhRvD..88j3510Z}.  In this model, the reconstructed velocity becomes unbiased when the smoothing term matches this damping effect.}.  As the galaxy number density decreases, $\alpha = 1$ can be retained by smoothing the galaxy field \citep{2021MNRAS.502.3723H}.  Velocity field bias as a function of smoothing is also discussed by \cite{2000ApJ...537..537B}.

\section{Numerical simulation tests}
\label{sec:method}

In this section we present the numerical simulations we generated to validate the velocity correlation models presented in Sec.\ref{sec:model}, and to compare the reconstruction-and-scaling and power-spectrum methods for measuring the growth rate.  We analysed two scenarios: a flat-sky case, in which we formed the dataset using a velocity component along a single axis; and a curved-sky case, in which our observable is the radial velocity relative to an observer.  These two scenarios allow us to construct convenient tests of reconstruction using the Fourier basis, and the spherical Fourier-Bessel basis, respectively.

\begin{figure*}
\centering
\includegraphics[width=1.8\columnwidth]{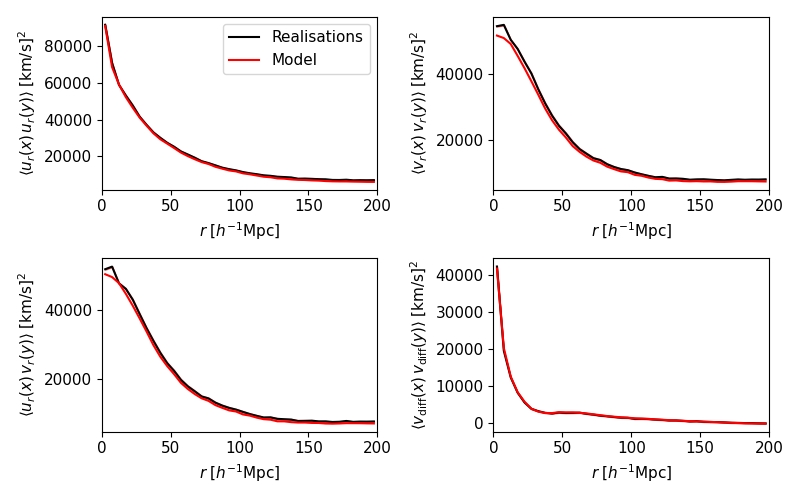}
\caption{Validation of the velocity correlation models described in Sec.\ref{sec:model} for the SFB-based reconstruction.  We generated many Gaussian realisations of the density and radial velocity fields, applied the SFB reconstruction method, and sampled the underlying and reconstructed velocity fields at a fixed set of locations.  We then measured the auto-correlation of the underlying velocity field $\langle \mathrm{u}_r(\mathbf{x}) \, \mathrm{u}_r(\mathbf{y}) \rangle$, the reconstructed velocity field $\langle \mathrm{v}_r(\mathbf{x}) \, \mathrm{v}_r(\mathbf{y}) \rangle$ and their cross-correlation $\langle \mathrm{u}_r(\mathbf{x}) \, \mathrm{v}_r(\mathbf{y}) \rangle$, by averaging over the realisations.  For convenience of display, we binned the results by separation $r = |\mathbf{x} - \mathbf{y}|$, where the measurements from the realisations (the black line) are compared to the predictions of the correlation models (the red line).  The final panel displays a similar comparison for the correlation of the velocity differences defined by Eq.\ref{eq:vdiff}.  The correlation models agree well with the measurements in all cases.}
\label{fig:velcorr}
\end{figure*}

\subsection{Simulation configuration}
\label{sec:sim}

Since our correlation models assume the density and velocity fields are Gaussian random fields, it was convenient for us to generate our numerical simulations as Gaussian realisations\footnote{We also studied lognormal realisations, which produced similar qualitative results to those presented here, but did not allow a precise validation of our velocity correlation model expressions due to the differing relationship between the density and velocity fields.}.  We generated Gaussian density fields from a fiducial model power spectrum in a cubic box of side length $600 \, h^{-1} \mathrm{Mpc}$ with a grid size of $128^3$.  We adopted the same fiducial cosmological model as \cite{2023MNRAS.526..337T}: matter density $\Omega_m = 0.273$, baryon density $\Omega_b = 0.0456$, Hubble parameter $h = 0.705$, normalisation $\sigma_8 = 0.812$ and spectral index $n_s = 0.96$, generating the model power spectrum using the {\tt CAMB} and {\tt halofit} prescriptions \citep{2000ApJ...538..473L, 2003MNRAS.341.1311S}.  We generated the model at $z=0$, and hence adopted fiducial parameters $f = 0.49$ and $b = 1$.  We then used Eq.\ref{eq:ui} to convert the density field to the components of the underlying velocity field in the linear-theory approximation.  Afterwards, we added noise to each grid cell of the density field equivalent to the Poisson noise of $N_g = 10^5$ objects uniformly distributed in the cube.  We are creating here a simple, uniform test configuration rather than seeking to model a realistic density-field sample, which would feature a galaxy number density varying with redshift and peaking at significantly higher values than our scenario \citep[e.g.,][]{2011MNRAS.416.2840L}.  However, the number of objects in our test dataset is comparable to current density-field samples used for this purpose.

In the flat-sky scenario, we analysed the density field within the whole cube, and assigned the velocity variable as the $x$-component of velocity.  In the curved-sky scenario, we placed an observer at the centre of the cube, cut the density field within a sphere of radius $R_g = 300 \, h^{-1} \mathrm{Mpc}$ centred on the observer, and assigned the velocity variable as the radial velocity relative to the observer.  In both scenarios we created the velocity sample by selecting $N_v = 10^4$ random positions within a sphere of radius $R_v$ with origin at the centre of the cube, assigning the value of the velocity field at each tracer's location (hence, we neglected clustering of the velocity tracers).  We considered different choices $R_v = \{ 100, 150, 200, 250, 300 \} \, h^{-1} \mathrm{Mpc}$, since we aimed to explore the dependence of the growth rate errors on the volume covered by the velocity sample.  In both scenarios we added a constant measurement noise $\epsilon_\mathrm{u} = 1000 \, \mathrm{km} \, \mathrm{s}^{-1}$ to the velocity values\footnote{The measurement noise for a realistic peculiar velocity dataset would grow in proportion to distance -- our intention is to create a simplified test configuration to validate our correlation models.}.

In the flat-sky scenario, we reconstructed the model velocity field from the noisy overdensity field in the cube by applying Eq.\ref{eq:ui}, after smoothing the density field with a Gaussian of standard deviation $\lambda = 10 \, h^{-1} \mathrm{Mpc}$ (noting that we obtain comparable results for smaller smoothing scales).  We used this reconstructed field to assign the model value of the velocity at each tracer's location.  In the curved-sky scenario, we determined the SFB coefficients from the noisy density field within the sphere by applying Eq.\ref{eq:denstosfb}.  We adopted a resolution $\{ n_\mathrm{max} = 9, \ell_\mathrm{max} = 24 \}$ which corresponds to preserving all SFB modes with $k_{n\ell} < 0.1 \, h^{-1} \mathrm{Mpc}$ (our results do not qualitatively depend on this choice, as long as use the same resolution when evaluating the covariance model).  We then reconstructed the model velocity at the location of each tracer by applying Eq.\ref{eq:vrecsfb}.  We created 400 different realisations of each of these mock datasets.

\subsection{Testing the velocity correlation models}

Before applying the different growth rate analysis methodologies to these mock datasets, we used the numerical simulations to test the velocity correlation models.  In order to create a precise test we fixed the positions of the velocity tracers for these purposes, and for each realisation we re-generated the Gaussian density field without changing the sampling locations.  We then measured the average value of the observed and model velocity correlations $\langle \mathrm{u}(\mathbf{x}) \, \mathrm{u}(\mathbf{y}) \rangle$, $\langle \mathrm{v}(\mathbf{x}) \, \mathrm{v}(\mathbf{y}) \rangle$ and $\langle \mathrm{u}(\mathbf{x}) \, \mathrm{v}(\mathbf{y}) \rangle$ as a function of separation between the unique pairs of velocity tracers.

The results are shown in Fig.\ref{fig:velcorr}, using the example of the SFB-based reconstruction in the curved-sky scenario.  Here, we computed the model correlations between the pairs of tracers using Eq.\ref{eq:ururcov}, \ref{eq:vrvrcovsfb} and \ref{eq:urvrcovsfb}.  The first three panels of Fig.\ref{fig:velcorr} show that the models accurately trace the numerical results\footnote{The small deviations between the models and measurements in the cases of $\langle \mathrm{v}_r(\mathbf{x}) \, \mathrm{v}_r(\mathbf{y}) \rangle$ and $\langle \mathrm{u}_r(\mathbf{x}) \, \mathrm{v}_r(\mathbf{y}) \rangle$ arise because, for reasons of computational speed, we evaluated the spherical harmonic coefficients in Eq.\ref{eq:vrecsfb} on a {\tt healpix} grid rather than at the precise angular positions of each tracer.  However, these small differences have no material effect on the ensuing analysis.}.  For the Fourier-based reconstruction in the flat-sky scenario, the correlation models and measurements also agree within the (small) numerical noise.

\subsection{Application of the reconstruction-and-scaling method}

In this section we discuss fitting the growth rate for each of our Gaussian realisations using the reconstruction-and-scaling method, in which we compare the measured galaxy velocities with the reconstructed velocity field values at the locations of these galaxies.  Given the velocity field is reconstructed assuming a fiducial value of $\beta = f/b$ and is linearly proportional to this quantity, the value of $\beta$ may be measured by performing a maximum-likelihood analysis scaling the reconstructed velocity values by $\beta/\beta_\mathrm{fid}$.  Based on the discussion in Sec.\ref{sec:bias}, we note that the reconstructed velocity field may be biased owing to the statistics of the velocity correlations.  Therefore to obtain an unbiased estimate of $\beta$, we must correct the model velocity field for this bias.

To apply the reconstruction-and-scaling method we therefore minimize as a function of $\beta$ the chi-squared statistic,
\begin{equation}
    \chi^2 = \mathbf{w}^T \, \mathbf{C}^{-1} \mathbf{w} ,
\label{eq:chisqscaling}
\end{equation}
where $\mathbf{w}$ is the vector of velocity data and model differences for each galaxy, and $\mathbf{C}$ is the covariance matrix associated with those differences.  For the $n^{\mathrm{th}}$ galaxy at location $\mathbf{x}_n$,
\begin{equation}
    w_n = \mathrm{u}(\mathbf{x}_n) - \alpha(\mathbf{x}_n) \frac{\beta}{\beta_\mathrm{fid}} \mathrm{v}(\mathbf{x}_n) ,
\label{eq:vdiff}
\end{equation}
where $\mathrm{u}(\mathbf{x})$ represents the measured velocity, $\mathrm{v}(\mathbf{x})$ is the reconstructed velocity, and $\alpha(\mathbf{x}) = r(\mathbf{x}) \, \sigma_\mathrm{u}(\mathbf{x}) / \sigma_\mathrm{v}(\mathbf{x})$ removes the statistical bias in the velocity field.

The elements of the covariance matrix $\mathbf{C}$ can be determined using the velocity correlations derived in Sec.\ref{sec:model}.  If $\beta$ is reasonably well-determined, it is a good approximation to evaluate the covariance assuming $\beta = \beta_\mathrm{fid}$.  In this case Eq.\ref{eq:vdiff} implies,
\begin{equation}
\begin{split}
    &C_{mn} = \langle \mathrm{u}(\mathbf{x}_m) \, \mathrm{u}(\mathbf{x}_n) \rangle - \alpha \langle \mathrm{v}(\mathbf{x}_m) \, \mathrm{u}(\mathbf{x}_n) \rangle \\ &- \alpha \langle \mathrm{u}(\mathbf{x}_m) \, \mathrm{v}(\mathbf{x}_n) \rangle + \alpha^2 \langle \mathrm{v}(\mathbf{x}_m) \, \mathrm{v}(\mathbf{x}_n) \rangle + \epsilon_u^2 \, \delta^K_{mn} ,
\end{split}
\label{eq:covdiff}
\end{equation}
where the correlations involving the model velocity depend on the reconstruction method utilised.  In the flat-sky scenario we use the $x$-component of velocities in this expression, and in the curved-sky scenario we use the radial velocity correlations.  For the diagonal elements, assuming the Fourier-based velocity reconstruction method, we can combine Eq.\ref{eq:siguu}, Eq.\ref{eq:sigvv} and Eq.\ref{eq:siguv} to show,
\begin{equation}
\begin{split}
    C_{nn} = \frac{a^2 H^2 f^2}{6\pi^2} \int dk & \left\{ P_m(k) \, [ 1 - \alpha D(k) ]^2 \right. \\ &+ \left. \frac{D^2(k)}{n_0} \right\} + \epsilon_u^2 ,
\end{split}
\end{equation}
and the off-diagonal elements include both these terms and the same additional factors as shown in Eq.\ref{eq:uiujcov} and Eq.\ref{eq:ururcov}.

These off-diagonal covariance terms are typically not included in the likelihood expressions of reconstruction-and-scaling approaches in the literature (for example: Eq.22 of \cite{2015MNRAS.450..317C}, Eq.26 of \cite{2020MNRAS.497.1275S}, Eq.9 of \cite{2020MNRAS.498.2703B}).  We acknowledge that our own analysis only represents a simplified test scenario and excludes many of the important astrophysical effects included in these aforementioned studies.  Hence, we cannot accurately assess the implications of these off-diagonal terms for existing observational studies.  However, we do note that in our test scenario, the covariance will only be well-described by noise alone ($C_{mn} = \epsilon_u^2 \, \delta^K_{mn}$) in the absence of velocity bias ($\alpha = 1$), smoothing ($D = 1$), a survey selection function ($W(\mathbf{x}_n) = n_0$) and shot noise ($n_0 \rightarrow \infty$).  Conversely, the presence of any of these factors may induce residual covariances between the velocity differences beyond those caused by noise.

We used our numerical simulations to validate the model of Eq.\ref{eq:covdiff}, which describes the correlation of velocity differences measured by Eq.\ref{eq:vdiff}.  The final panel of Fig.\ref{fig:velcorr} displays the average correlation $C_{mn} = \langle w(\mathbf{x}_m) \, w(\mathbf{x}_n) \rangle$ as a function of separation for the curved-sky scenario, showing excellent agreement between the numerical simulations and theoretical model.

We used the reconstruction-and-scaling method to determine the best-fitting value and error in $\beta$ for each of the 400 Gaussian realisations, by using the $\chi^2$ statistic defined in Eq.\ref{eq:chisqscaling} as a likelihood.  We repeated this analysis for the different choices of the radial extent of the velocity sample, $R_v = \{ 100, 150, 200, 250, 300 \} \, h^{-1} \mathrm{Mpc}$.  We compared results using the full covariance matrix of Eq.\ref{eq:covdiff} and a covariance matrix including only noise, $C_{mn} = \epsilon_u^2 \, \delta^K_{mn}$.

\subsection{Application of the power-spectrum method}

\begin{figure*}
\centering
\includegraphics[width=2\columnwidth]{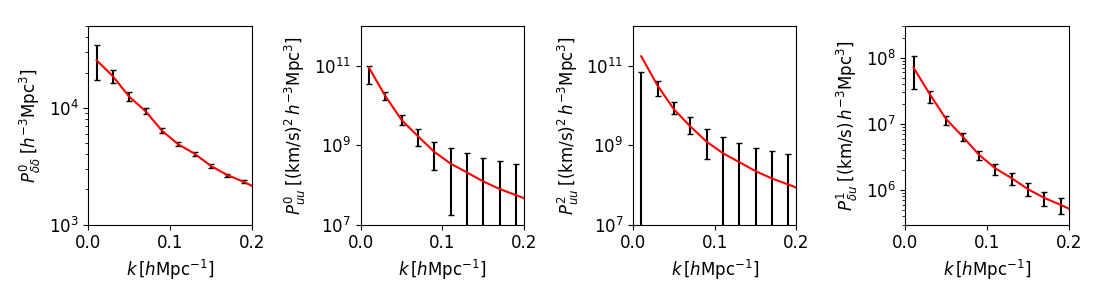}
\caption{The mean and standard deviation of the power spectrum measurements across our numerical realisations (black data points and errors), compared with the theoretical models (red lines).  The panels display the four power spectra which contain the information about the density and velocity fields (in the absence of redshift-space distortions): the monopole of the density power spectrum $P^0_{\delta\delta}$, the monopole and quadrupole of the velocity power spectrum $\left\{ P^0_{uu}, P^2_{uu} \right\}$, and the dipole of the cross-power spectrum, $P^1_{\delta u}$.}
\label{fig:pkmeas}
\end{figure*}

As a comparison with the reconstruction-and-scaling method, we performed a fit for $\beta$ using the 2-point power spectra of the velocity and density fields.  In linear theory, neglecting redshift-space distortions, the joint information in the density and velocity fields is contained in four power spectra: the monopole of the density power spectrum $P^0_{\delta\delta}$, the monopole and quadrupole of the velocity power spectrum $\left\{ P^0_{uu}, P^2_{uu} \right\}$, and the dipole of the cross-power spectrum, $P^1_{\delta u}$ \citep[e.g.,][]{2019MNRAS.487.5209H}.  If redshift-space distortions were included, the quadrupole of the density power spectrum $P^2_{\delta\delta}$ would also be significant.

We measured these power spectra from our numerical simulations using Fast Fourier Transform (FFT) methods.  As described in Sec.\ref{sec:sim}, in the flat-sky scenario we utilised the full overdensity grid of the realisation over the cube of volume $V_\mathrm{box}$, which includes Poisson noise equivalent to $N_g$ density tracers.  We formed the observable velocity field by binning onto the same grid the discretely-sampled $x$-components of velocities at the locations of the $N_v$ tracers, which are distributed within a sphere of volume $V_v = \frac{4}{3} \pi R_v^3$.  We then normalised the velocity grid by dividing by $N_v$.  With these conventions, we could estimate the auto- and cross-power spectra as a function of wavenumber $\mathbf{k}$ as,
\begin{equation}
\begin{split}
    \hat{P}_{\delta\delta}(\mathbf{k}) &= \left( |\tilde{\delta}(\mathbf{k})|^2 \,  - \frac{1}{N_g} \right) \, V_\mathrm{box} , \\
    \hat{P}_{uu}(\mathbf{k}) &= \left( |\tilde{u}(\mathbf{k})|^2 - \frac{\sigma_u^2}{N_v} \right) \, V_\mathrm{box} , \\
    \hat{P}_{\delta u}(\mathbf{k}) &= \mathrm{Im} \{ \tilde{\delta}(\mathbf{k}) \, \tilde{u}^*(\mathbf{k}) \} \, V_\mathrm{box} .
\end{split}
\end{equation}
In these estimators, $\tilde{\delta}(\mathbf{k})$ is the Fourier transform of the overdensity grid, $\tilde{u}(\mathbf{k})$ is the Fourier transform of the normalised velocity grid, and $\sigma_u^2$ is the variance of the velocity values.  Finally, we scaled the velocity auto-power spectrum measurements by a factor of $V_\mathrm{box}/V_v$ to produce an unbiased estimate allowing for the partial coverage of the velocity sample within the box \citep[e.g.,][]{2013MNRAS.436.3089B}.

The power spectrum multipole measurements follow from these estimates as,
\begin{equation}
\hat{P}_X^\ell(k) = (2\ell+1) \int \frac{d\Omega_k}{4\pi} \, \hat{P}_X(\mathbf{k}) \, L_\ell(\mu_k) ,
\label{eq:pkestmult}
\end{equation}
where $X = \{ \delta\delta , uu, \delta u \}$, $\mu_k = k_x/k$ is the cosine of the angle of the wavevector to the line-of-sight (which is fixed in this scenario), and we averaged over all modes within a spherical shell $k = |\mathbf{k}|$, where we take the bin width as $\Delta k = 0.02 \, h \, \mathrm{Mpc}^{-1}$.  The theoretical models for the power spectra can be deduced from Eq.\ref{eq:uk} as,
\begin{equation}
\begin{split}
\langle \hat{P}_{\delta\delta}(\mathbf{k}) \rangle &= b^2 \, P_m(k) , \\
\langle \hat{P}_{uu}(\mathbf{k}) \rangle &= \frac{a^2 H^2 \beta^2 \, k_x^2 \, b^2 \, P_m(k)}{k^4} , \\
\langle \hat{P}_{\delta u}(\mathbf{k}) \rangle &= \frac{a H \beta \, k_x \, b^2 \, P_m(k)}{k^2} ,
\end{split}
\label{eq:pkmod}
\end{equation}
where we have explicitly expressed the models in terms of $\beta = f/b$, enabling us to perform a joint fit for $\{ \beta, b \}$.  Substituting these models into Eq.\ref{eq:pkestmult}, we find for the power spectrum multipoles,
\begin{equation}
\begin{split}
    \langle \hat{P}^0_{\delta\delta}(k) \rangle &= b^2 P_m(k) , \\
    \langle \hat{P}^0_{uu}(k) \rangle , \langle \hat{P}^2_{uu}(k) \rangle &= \left\{ \frac{1}{3} , \frac{2}{3} \right\} \frac{a^2H^2\beta^2}{k^2} \, b^2 P_m(k) , \\
    \langle \hat{P}^1_{\delta u}(k) \rangle &= \frac{aH\beta}{k} \, b^2 P_m(k) .
\end{split}
\end{equation}
We convolved the model power spectra with the window functions of each tracer.

For the curved-sky scenario we analysed the density field within a spherical volume with radius $R_g = 300 \, h^{-1} \mathrm{Mpc}$, and used the radial velocity as our observable.  Given that the direction of the line-of-sight now changes across the volume, we followed the method of \cite{2015MNRAS.453L..11B} to estimate the quadrupole and dipole power spectra as,
\begin{equation}
\begin{split}
    \hat{P}^2_{uu}(k) &= 5 \int \frac{d\Omega_k}{4\pi} A_{0u}(\mathbf{k}) \left[ 3 A^*_{2u}(\mathbf{k}) - A^*_{0u}(\mathbf{k}) \right] , \\
    \hat{P}^1_{\delta u}(k) &= 3 \int \frac{d\Omega_k}{4\pi} A_{0\delta}(\mathbf{k}) \, A^*_{1u}(\mathbf{k}) ,
\end{split}
\label{eq:pkcurvest}
\end{equation}
where the monopole estimators are unchanged.  Following the notation of \cite{2015MNRAS.453L..11B} the terms in Eq.\ref{eq:pkcurvest} are,
\begin{equation}
    A_{n f}(\mathbf{k}) = \frac{1}{V_\mathrm{box}} \int d^3\mathbf{x} \, f(\mathbf{x}) \, \left( \mathbf{\hat{k}} \cdot \mathbf{\hat{x}} \right)^n \, e^{i\mathbf{k} \cdot \mathbf{x}} ,
\end{equation}
where $f(\mathbf{x}) = \delta(\mathbf{x})$ for the density field, and $f(\mathbf{x}) = u_r(\mathbf{x})/N_v$ for the radial velocity field.  These estimators may be evaluated using FFT methods by writing $\mathbf{\hat{k}} \cdot \mathbf{\hat{x}} = \sum_i k_i x_i / kx$ and expanding the expressions \citep{2015MNRAS.453L..11B}.  Again, we scale the power spectrum measurements by the coverage fraction of each sample inside the box.

Whilst analytical methods are available to evaluate the covariance matrices of power spectrum multipole measurements \citep{2016MNRAS.457.1577G, 2018MNRAS.479.5168B}, it was convenient for us to determine the covariance of the joint data vector $\{ P^0_{\delta\delta} , P^0_{uu} , P^2_{uu} , P^1_{\delta u} \}$ by averaging over the ensemble of numerical realisations. Fig.\ref{fig:pkmeas} displays the mean of the power spectrum measurements across the realisations for the curved-sky scenario, demonstrating good agreement with our models.

We used the power-spectrum method to determine the best-fitting value and error in $\beta$ for each of the 400 Gaussian realisations, by performing a maximum-likelihood for $\{ \beta, b \}$ and marginalising over the bias parameter.  We fit to a range of scales $k < 0.2 \, h \mathrm{Mpc}^{-1}$, although our $\beta$ measurements are not sensitive to this choice, given that the signal-to-noise contained in the velocity power is dominated by low-$k$ modes.  We repeated this analysis for the different choices of the radial extent of the velocity sample, $R_v = \{ 100, 150, 200, 250, 300 \} \, h^{-1} \mathrm{Mpc}$.  In all cases we found that, as expected for this simulation approach, the average minimum $\chi^2$ of the fits was equal to the number of degrees of freedom, and the average best-fitting value of $\beta$ was equal to the fiducial input value.

\vspace{12pt}

\section{Results}
\label{sec:results}

In this section we compare the error in determining the value of $\beta$ based on the different techniques, inferred from the ensemble of numerical simulations, as a function of the radial extent of the velocity sample.  We distinguish two methods for quantifying the error in $\beta$. First, we calculate the ``mean error'' in $\beta$ across the realisations, which reflects the average error that would be quoted by an experimenter.  Second, we determine the ``standard deviation in the best fits'' across the realisations, which quantifies the true accuracy of the method across an ensemble of universes.  Where these error estimates agree, we can conclude that the error determination is unbiased.

\begin{figure}
\centering
\includegraphics[width=\columnwidth]{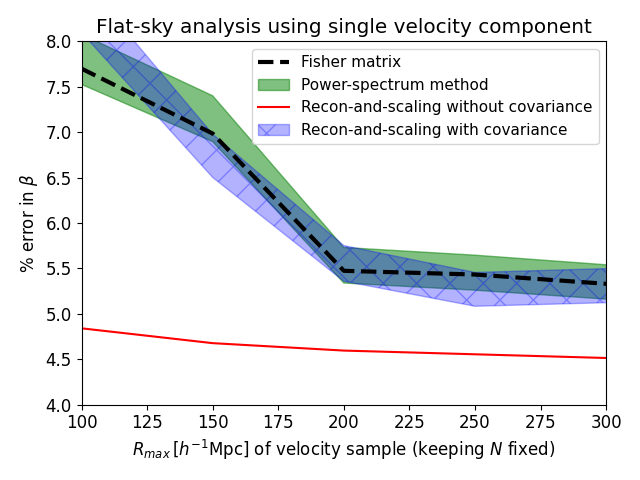}
\caption{The percentage error in the recovered $\beta$ values for the flat-sky analysis, as a function of the radial extent of the velocity sample.  We compare the error forecast by a Fisher matrix with that obtained by the power-spectrum method, the reconstruction-and-scaling method using the full covariance, and the reconstruction-and-scaling method with noise-only errors.  The bands show the ``error in the error'' deduced from the number of realisations assuming Gaussian statistics, which have width $\sigma_\beta/\sqrt{2 N_\mathrm{real}}$.}
\label{fig:betaerrflat}
\end{figure}

\begin{figure}
\centering
\includegraphics[width=\columnwidth]{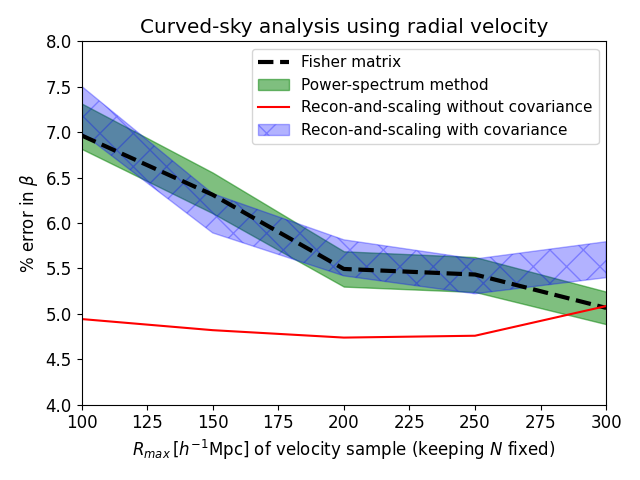}
\caption{The percentage error in the recovered $\beta$ values for the curved-sky analysis, comparing the different methods in the same style as Fig.\ref{fig:betaerrflat}.}
\label{fig:betaerrcurv}
\end{figure}

Fig.\ref{fig:betaerrflat} displays the mean errors for the flat-sky scenario, and Fig.\ref{fig:betaerrcurv} contains the same information for the curved-sky scenario.  We compare results for the power-spectrum method, the reconstruction-and-scaling method fit using the full covariance of Eq.\ref{eq:covdiff}, and the reconstruction-and-scaling method fit using the noise-only ``diagonal'' errors.  For the power-spectrum method and reconstruction-and-scaling method using the full covariance, the mean error is very similar to the standard deviation in the best fits.  However, for the reconstruction-and-scaling method fit using the noise-only errors, the standard deviation in the best fits is significantly larger than the mean error, indicating that the error determined by this method is under-estimated.

As a further point of comparison, we forecast the expected error in $\beta$ using a Fisher matrix analysis based on the four power spectrum models $\mathbf{m} = \{ P^0_{\delta\delta}(k) , P^0_{uu}(k) , P^2_{uu}(k) , P^1_{\delta u}(k) \}$ and their joint covariance $\mathbf{C}$.  To do this we evaluated the Fisher matrix,
\begin{equation}
    \mathbf{F}_{ij} = \frac{\partial\mathbf{m}^T}{\partial p_i} \, \mathbf{C}^{-1} \, \frac{\partial\mathbf{m}}{\partial p_j} ,
\end{equation}
as a function of the parameters $p_i = \{ \beta, b \}$.  The forecast error in $\beta$ is then $\sigma_\beta = \sqrt{ \left( \mathbf{F}^{-1} \right)_{11} }$.  We find that the Fisher matrix forecast agrees well with the error determined by both the power-spectrum method, and the reconstruction-and-scaling method using the full covariance.  We note in Fig.\ref{fig:betaerrcurv} that there is a small numerical systematic in the curved-sky reconstruction-and-scaling method results when $R_v = 300 \, h^{-1}\mathrm{Mpc}$, which matches the radius at which the SFB boundary conditions are applied\footnote{This numerical issue could be addressed by changing the boundary conditions, but exploring these effects was unimportant for the interpretation of the results.}.

It is notable from Fig.\ref{fig:betaerrflat} and Fig.\ref{fig:betaerrcurv} that the fractional error in $\beta$ increases as the radial extent of the velocity sample reduces, when this error is estimated in an unbiased fashion.  For the reconstruction-and-scaling method assuming a noise-only covariance, the inferred fractional error in $\beta$ does not change with the volume of the velocity sample.

We can interpret these results as follows: in the power spectrum analysis approach, a different sample variance is associated with the density and velocity tracers, depending on their respective window functions.  If the velocity sample reduces in volume, its sample variance would correspondingly increase, enlarging the final errors in $\beta$.  In the reconstruction-and-scaling method, the equivalent effect is that when the velocity tracers are distributed over a smaller volume, the ``average correlation'' between pairs of them increases, reducing their ``statistical power''.  These effects appear in the covariance between the tracers -- hence, when this covariance is neglected, the error in $\beta$ appears to be independent of volume.  However, this is an artefact of the adopted correlation model.

\section{Conclusions}
\label{sec:conc}

In this study we have derived expressions for the analytical covariance of the velocity field reconstructed from a set of density tracers, considering both a Fourier basis within a cuboid, and a spherical Fourier-Bessel basis within a curved-sky observational volume.  We have applied these results to explore the errors in the growth rate parameter that arise when fitting an observational peculiar velocity dataset using the reconstruction-and-scaling method, using an ensemble of Gaussian realisations designed to model key features of this scenario.  We compared these parameter constraints to those resulting from a joint analysis of the density and velocity auto- and cross-power spectra.

We conclude that the power-spectrum method and reconstruction-and-scaling method including the full covariance both produce unbiased estimates of the error in the growth rate parameter, which approximately agree with the Fisher matrix forecast.  However, the reconstruction-and-scaling method may underestimate the true error if the fit is carried out using only diagonal noise terms.  This issue occurs because the reconstructed velocity field contains sample variance, which must be propagated into the resulting growth rate error.  We treat this finding as a potential explanation of the difference between the growth rate errors previously reported by the two methods in the literature.

Our numerical simulations are designed as a simplified and representative test scenario, and purposefully exclude several properties of realistic datasets such as survey window functions, distance-dependence of measurement errors, redshift-space distortions and non-linear galaxy bias.  Future work will study these aspects using more realistic simulations and real datasets.

\section*{Acknowledgements}

We thank the anonymous referees for providing useful comments.  We are grateful for useful conversations with Khaled Said, Matthew Colless and Paula Boubel, which sparked our curiosity to begin this project.  We also thank Mike Hudson and Khaled Said for providing valuable feedback on a draft of this paper.  We acknowledge financial support from Australian Research Council Discovery Project DP220101610.

\section*{Data availability}

The data underlying this article will be shared on reasonable request to the corresponding author.

\newpage

\bibliographystyle{mnras}
\bibliography{pv_errors}

\appendix

\section{Correlation of reconstructed velocities with a window function}
\label{sec:approxcoeff}

In this Appendix we provide the exact expression for the correlation of velocity field components between two positions $\mathbf{x}$ and $\mathbf{y}$, reconstructed using the Fourier method, in the presence of a window function $W(\mathbf{x})$.  After substituting Eq.\ref{eq:dgdg} in Eq.\ref{eq:uk} we find,
\begin{equation}
    \langle \mathrm{v}_i(\mathbf{x}) \, \mathrm{v}_j(\mathbf{y}) \rangle = \frac{a^2 H^2 f^2}{n_0^2} \left[ \int \frac{d^3\mathbf{z}}{V_\mathrm{box}} \, W(\mathbf{z}) \, g_i(\mathbf{z} - \mathbf{x}) \, g_j^*(\mathbf{z} - \mathbf{y}) + \int \frac{d^3\mathbf{k}}{(2\pi)^3} P_m(\mathbf{k}) \, f_i(\mathbf{x},\mathbf{k}) \, f^*_j(\mathbf{y},\mathbf{k}) \right] ,
\end{equation}
where we have defined,
\begin{equation}
\begin{split}
  f_i(\mathbf{x},\mathbf{k}) &= \int \frac{d^3\mathbf{k}'}{(2\pi)^3} \, \frac{k_i'}{k'^2} \, \tilde{W}(\mathbf{k}'-\mathbf{k}) \, e^{-i\mathbf{k}' \cdot \mathbf{x}} , \\
  g_i(\mathbf{x}) &= \int \frac{d^3\mathbf{k}'}{(2\pi)^3} \, \frac{k_i'}{k'^2} \, e^{-i \mathbf{k}' \cdot \mathbf{x}} .
\end{split}
\end{equation}
These expressions are cumbersome to evaluate, but an approximation is available by assuming $k_i'/k'^2$ is more slowly varying with $\mathbf{k}'$ than $\tilde{W}(\mathbf{k}'-\mathbf{k})$.  This approximation is equivalent to supposing that the window function is slowly-varying on the correlation scale in question.  Simplifying the expressions using this approximation and evaluating the covariance of the reconstructed radial velocities from the components, we find that $\langle \mathrm{v}(\mathbf{x}) \, \mathrm{v}(\mathbf{y}) \rangle$ can be written in the same form as Eq.\ref{eq:uiujcov} and Eq.\ref{eq:ururcov} after making the replacement,
\begin{equation}
    P_m(k) \rightarrow  \left[ \frac{P_m(k) \, W(\mathbf{x}) \, W(\mathbf{y})}{n_0^2} + \frac{W(\mathbf{x}) + W(\mathbf{y})}{2 \, n_0^2} \right] D^2(k) .
\end{equation}
This approximation preserves the appropriate symmetry under the interchange $\mathbf{x} \leftrightarrow \mathbf{y}$.  In this approximation, the variance of the reconstructed velocity at a point is given by,
\begin{equation}
    \langle \mathrm{v}^2(\mathbf{x}) \rangle = \frac{a^2 H^2 f^2}{6\pi^2} \int dk \, \left[ P_m(k) \frac{W^2(\mathbf{x})}{n_0^2} \right. \left. + \frac{W(\mathbf{x})}{n_0^2} \right] D^2(k) ,
\end{equation}
which may be compared with Eq.\ref{eq:sigvv}, which neglects the window function.  Similarly, the cross-correlation between the underlying and reconstructed velocities may be approximated by making the replacement in Eq.\ref{eq:uiujcov} and Eq.\ref{eq:ururcov}, $P_m(k) \rightarrow P_m(k) \, D(k) \, W(\mathbf{x})/n_0$, where $\mathbf{x}$ is the location of the reconstructed velocity.  In this case, the covariance of the velocity fields at a point is given by,
\begin{equation}
    \langle \mathrm{u}(\mathbf{x}) \, \mathrm{v}(\mathbf{x}) \rangle = \frac{a^2 H^2 f^2}{6\pi^2} \int dk \, P_m(k) \, D(k) \, \frac{W(\mathbf{x})}{n_0} ,
\end{equation}
which may be compared with Eq.\ref{eq:siguv}.

\section{Agreement between SFB and underlying velocity correlation in the limit $R \rightarrow \infty$}
\label{sec:sfbtocont}

As a consistency check of the correlation between velocities reconstructed using the SFB method, we show in this Appendix that Eq.\ref{eq:vrvrcovsfb} reduces to Eq.\ref{eq:ururcov} in the limit that $R \rightarrow \infty$.  In this limit we note that,
\begin{equation}
\begin{split}
    W_{n \ell}(k) &\rightarrow \frac{\pi}{2 k^2_{n \ell}} \, \delta_D(k - k_{n \ell}) , \\
    C_{n n' \ell} &\rightarrow \frac{\pi^2}{4} \, k^{-2}_{n \ell} \, P(k_{n \ell}) \, \delta^K_{n n'} , \\
    c_{n \ell} &\rightarrow \frac{2}{\pi} k^2_{n \ell} ,
\end{split}
\end{equation}
where $k_{n \ell} \rightarrow k$ and $\sum_n \rightarrow \int dk$.  With these substitutions Eq.\ref{eq:vrvrcovsfb} becomes,
\begin{equation}
    \langle \mathrm{v}_r(\mathbf{x}) \, \mathrm{v}_r(\mathbf{y}) \rangle = a^2 H^2 f^2 \int_0^\infty \frac{dk}{2\pi^2} \, P_m(k) \left[ \sum_\ell (2\ell+1) \, j'_\ell(kx) \, j'_\ell(ky) \, L_\ell(\hat{\mathbf{x}} \cdot \hat{\mathbf{y}}) \right] .
\end{equation}
In order to reduce this relation to the form of Eq.\ref{eq:ururcov}, we employ Gegenbauer's addition theorem,
\begin{equation}
    j_0(w) = \sum_\ell (2\ell + 1) \, j_\ell (u) \, j_\ell(v) \, L_\ell(\cos{\alpha}) ,
\label{eq:gegenbauer}
\end{equation}
where $\{ u, v, w \}$ form three sides of a triangle where the angles opposite these sides are $\{ \gamma, \beta, \alpha \}$, respectively.  By evaluating the quantity $\partial^2 j_0(w)/\partial u \, \partial v$, Eq.\ref{eq:gegenbauer} becomes,
\begin{equation}
    \frac{j_1(w)}{w} \cos{\alpha} + j_2(w) \cos{\beta} \cos{\gamma} = \sum_\ell (2\ell + 1) \, j'_\ell (u) \, j'_\ell(v) \, L_\ell(\cos{\alpha}) .
\end{equation}
Identifying the three sides $\{ u, v, w \}$ with the position and separation vectors $\{ \mathbf{x}, \mathbf{y}, \mathbf{r} \}$, we can then recover Eq.\ref{eq:ururcov} by using $\cos{\beta} = \hat{\mathbf{x}} \cdot \hat{\mathbf{r}}$ and $\cos{\gamma} = -\hat{\mathbf{y}} \cdot \hat{\mathbf{r}}$.

\end{document}